\documentstyle[multicol,aps,epsfig,floats]{revtex}

\begin{document}

\twocolumn[\hsize\textwidth\columnwidth\hsize\csname @twocolumnfalse\endcsname


\title{Transition temperature of ferromagnetic semiconductors:
a dynamical mean field study}
\author{A.\ Chattopadhyay$^{1,2}$, S. Das Sarma$^1$ and A.\ J.\ Millis$^2$}
\address{
$^1$ Department of Physics, University of Maryland\\
College Park, MD 20742\\
$^2$ Center for Materials Theory\\
Department of Physics and Astronomy, Rutgers University\\
Piscataway, NJ 08854\\
}
\date{\today}
\draft
\maketitle
\widetext
\begin{abstract}
\noindent We formulate a theory of doped magnetic semiconductors such as Ga$%
_{1-x}$Mn$_x$As which have attracted recent attention for their possible use
in spintronic applications. We solve the theory in the dynamical mean field
approximation to find the magnetic transition temperature $T_c$ as a
function of magnetic coupling strength $J$ and carrier density $n$. We find
that $T_c$ is determined by a subtle interplay between carrier density and
magnetic coupling.
\end{abstract}



 ]

\narrowtext


Diluted magnetic semiconductors have attracted much recent attention \cite
{Ohno98} for their potential use in spintronic devices. The prospect of
carrying out both information processing and storage on the same chip is an
exciting possibility. For applications it is desirable to find materials
which are ferromagnetic at as high a temperature as possible, so the
discovery \cite{Ohno96} of a ferromagnetic transition with $T_{c}$ as high
as 110K in MBE grown Ga$_{1-x}$Mn$_{x}$As, has inspired a great deal of
interest.  In addition to its potential technological significance, the
ferromagnetism of Ga$_{1-x}$Mn$_{x}$As is an important fundamental condensed
matter problem. The cause of ferromagnetism in Ga$_{1-x}$Mn$_{x}$As (and
similar materials, e.g In$_{1-x}$Mn$_{x}$As where $T_{c}\approx 25-30K$\cite
{Ohno96}) is controversial: the different proposed mechanisms\cite
{Matsukura98,Akai98,Dietl00,Konig00,Bhatt00,Litvinov01,Nagaev96}
do not qualitatively
agree with each other. In this paper we present a new theoretical approach
which allows calculation of magnetic transition temperatures (and other
properties) over a much wider temperature range than had previously been
possible and provides new insights into the factors controlling $T_{c}$.

It is well established that in III-V systems such as Ga$_{1-x}$Mn$_{x}$As,
the Mn ions go in substitutionally at the cation (Ga) sites and contribute
itinerant holes to the GaAs valence band. The experimental hole density $n$
is typically a small fraction ($10\%$ or so) of the Mn concentration perhaps
due to strong localization at As antisite defects so the Ga$_{1-x}$Mn$_{x}$%
As system could be considered partially compensated. The Mn ion has a half
filled d-shell and acts as a $S=5/2$ local moment; the itinerant carriers
are locally magnetically coupled to the Mn spins via an exchange coupling $J$
which may be either antiferromagnetic ($J>0$) (for electrons) or
ferromagnetic ($J<0$) (for holes : physical Ga$_{1-x}$Mn$_{x}$As case).\cite
{Sanvito01}.

It is generally accepted {\cite
{Ohno98,Ohno96,Matsukura98,Akai98,Dietl00,Konig00,Bhatt00,Litvinov01,Nagaev96,Sanvito01}
} that the magnetic semiconductors are described by a generalized Kondo
lattice model:
\begin{eqnarray}
H_{KL} &=&\sum_{i,j}J_{AF}(R_{i}-R_{j}){\bf S}_{i}\cdot {\bf S}%
_{j}\nonumber \\
&+&\sum_{u\alpha }\int d^{3}x\psi _{pu\sigma }^{+}(x)\left( \frac{-\nabla
^{2}}{2m_{u}}+V_{r}(x)\right) \psi _{pu\sigma }(x)  \nonumber \\
&+&\int d^{3}x\sum_{iu\alpha \beta }W(x-R_{i})\psi _{u\alpha }^{+}(x)\psi
_{u\alpha }(x)\nonumber \\
&+&J{\bf S}_{i}\cdot \psi _{u\alpha }^{+}(x)\sigma _{\alpha
\beta }\psi _{u\beta }(x)b^{3}\delta ^{3}(x-R_{i})  \label{HKL}
\end{eqnarray}
where $u$ labels the relevant bands of the semiconductor (the two hole bands
in hole-doped GaAs for example), $V_{r}$ is a potential arising from
randomness in the host lattice (e.g. As antisite defects), $R_{i}$ are the
positions of the Mn dopants, $W$ is the (presumably coulombic) potential
arising from the Mn dopant, $J_{AF}$ is a direct antiferromagnetic
exchange between Mn spins arising from other orbitals unrelated to the
doped holes, $J$ is the local exchange coupling between the spin of the Mn
and the spins of the semiconductor carriers. We normalize the $\delta $
function in the $J$ term to the volume $b^{3}=5.65{\AA}^{3}$ per GaAs unit.
The large (${\bf S}=5/2$) value of the Mn spin justifies treating the
spins classically, so that the partition function $Z$ may be determined by
finding the free energy $F(\{{\bf S}_{i}\})$ of holes in a fixed spin
configuration and then averaging over spin configurations, i.e.
\begin{equation}
Z=\int \{d{\bf S}_{i}\}e^{-\left( \sum_{i,j}J_{AF}(R_{i}-R_{j}){\bf S}%
_{i}\cdot {\bf S}_{j}+F(\{{\bf S}_{i}\})\right) /k_{B}T}  \label{Z}
\end{equation}
The key issue is therefore the evaluation of $F(\{{\bf S}_{i}\})$ . From Eq.~%
\ref{HKL}, we see that $F$ is the free energy of noninteracting carriers in
a spin dependent potential, which may have randomness both from the
distribution of Mn positions and from spin disorder. Further, as will be
shown explicitly below, the relevant temperatures are small compared to the
hole Fermi energy ($E_{F}$), so that $F$ is to a good approximation simply
the carrier ground state energy in the given spin configuration. The crucial
quantity governing $T_{c}$ is the {\it change} in $F$ as the spin
configuration goes from disordered to ordered. As we will show below, the
change in $F$ involves several competing effects not evident in previous
calculations of $T_{c}$ such as the static mean field theory\cite{Dietl00}.

To obtain more detailed information, we consider here the idealized Kondo
lattice model, in which $J_{AF},$ $V_{r}$ and $W$ in Eq.\ref{HKL} are
neglected. These may easily be included in our formalism, and will be
discussed in a future paper. We absorb the magnitude of the spin into the
definition of $J$ and study first the $T\rightarrow 0$ limit of the fully
polarized ferromagnetic state, ${\bf S}_{i}\parallel \widehat{z}$ at $T=0$.
In this state the carriers feel a spatially varying spin dependent potential
with mean strength $xJ$ per GaAs unit cell. It leads to a shift in the band
offset (upwards for one species and downwards for the other) proportional to
$xJ$. For values of $J$ less than a critical value $J_{c}$, the spin
dependent potential does not lead to any bound states. The wave functions
are scattering states and the  band offset is essentially $\pm xJ$. The
energy is given simply by filling the up and down bands up to the
appropriate chemical potential. The critical value $J_{c}$ corresponds to a
magnetic coupling strong enough to bind a hole of the appropriate spin to a
Mn site, causing a 'parallel-spin' impurity band to split off from the main
band. Sanvito et al.\cite{Sanvito01} used the local spin density
approximation (LSDA) and supercell methods to study  Ga$_{1-x}$Mn$_{x}$As with
a dilute but spatially ordered Mn lattice. The LSDA prediction for  $J$
was found to be very close to the critical value needed for impurity band
formation (Fig.11 of Ref.~\cite{Sanvito01}, shows a band shift,  within $15\%
$ of $\Delta E=10eV/$Mn, implying in our conventions $J=1eV$); experimental
estimates tend to be somewhat lower.

We now consider energetics of arbitrary spin configurations. In the small $J$
limit, straighforward perturbation theory shows that
\begin{equation}
\delta F=\frac{1}{2}\sum_{ij}J^{2}\chi (R_{i},R_{j}){\bf S}_{i}\cdot {\bf S}%
_{j}  \label{delF}
\end{equation}
with $\chi (R_{i},R_{j})$ the static spin susceptibility computed from Eq
\ref{HKL} with $J=0$. This is often referred to \cite
{Matsukura98,Akai98,Dietl00} as the 'RKKY' limit although strictly speaking
the term RKKY refers to the behavior of $\chi $ at distances long compared
to the spacing between carriers and Eq.~\ref{delF} applies even for spins
closer together than this distance. When Eq.~\ref{delF} applies, the
ordering wavevector is the one which maximizes $\chi$ and
$T_{c}\sim J^{2}$, which is
also the static mean-field result \cite{Dietl00}.
Note that for $n$ greater than a (numerically small)
critical value $n_c$, the maximum in  $\chi $
is at a nonzero wavevector, leading to a non-ferromagnetic ordered
state.

In the $J\rightarrow \infty $ limit, at all times each carrier is bound to
an Mn site with a binding energy proportional to $J$ and spin parallel to
the Mn spin on that site. The dependence of energy on spin configurations
arises because in the paramagnetic state some hopping processes are blocked
\cite{Amit00} and is therefore set by the impurity bandwidth which is never
large (because the Mn are dilute) and vanishes as $J\rightarrow \infty $
due to the contraction of the Bohr radius of the bound state. In this limit $%
T_{c}$ depends crucially on the impurity band filling. For a full impurity
band (one carrier per Mn) no low energy hopping processes are possible in
a fully polarized ferromagnetic state; the ground state for a nearly filled
impurity band is antiferromagentic or phase separated. The static mean field
theory \cite{Dietl00,Konig00} does not capture this physics at all,
predicting instead a $T_{c}\sim J^{2}$ for all $J$.

We now present a dynamical mean field theory which gives a reasonable
account of the small and intermediate $J$ regime as well as the crossover to
the 'impurity band' regime. It however does not adequately treat the band
narrowing arising from extreme wavefunction localization so  breaks down at
some $J>>J_{c}$. We model the GaAs:Mn system as a lattice of sites, which
are randomly nonmagnetic (with probability $1-x$) or magnetic (with
probability $x$). Standard arguments\cite{dmft} show that the relevant
physics may then be determined from the local (momentum-integrated) Green
function ${\bf G}_{loc}^{a,b}(\omega )=b^{3}\int \frac{d^{3}p}{\left( 2\pi
\right) ^{3}}(\omega -\Sigma _{\alpha \beta }^{a,b}(\omega )-\varepsilon
_{pa})^{-1}$. $G_{loc}$ is in general a matrix in spin and band (not shown)
indices  and depends on whether one is considering a magnetic ($a)$ or
non-magnetic ($b)$ site. Being a local function, it is the solution of a
local problem specified by the partition function $Z_{loc}=\int d{\bf Se}%
^{-S_{loc}}$ with action 
$S_{loc}=g_{0\alpha \beta }^{a}(\tau -\tau ^{\prime })c_{a\alpha }^{+}(\tau
)c_{a\beta }(\tau ^{\prime })+J{\bf S}\cdot \sum_{a\alpha \beta }c_{a\alpha
}^{+}(x){\bf \sigma }_{\alpha \beta }c_{a\beta }(x)$ on the $a$ (magnetic)
site and $S_{loc}=g_{0\alpha \beta }^{b}(\tau -\tau ^{\prime })
c_{a\alpha }^{+}(\tau)c_{a\beta }(\tau ^{\prime })$
on the non-magnetic ($b$) site.
The a-site mean field function $g_{0}^{a}$ can be written as $g_{0\alpha
\beta }^{a}=a_{0}+a_{1}{\widehat{m}}\cdot {\bf \sigma }_{\alpha \beta }$
with ${\widehat{m}}$ the magnetization direction and $a_{1}$ vanishing in
the paramagnetic state. It is specified by the condition that the local
Green function computed from $Z_{loc}$, namely $\delta \ln Z_{loc}/\delta
g_{0}^{a}=\left( g_{0}^{a}-\Sigma \right) ^{-1}$ is identical to the local
green function computed by performing the momentum integral using the same
self energy. The momentum integal requires an upper cutoff because the $%
p^{2}/2m$ dispersion given in Eq.~\ref{HKL} applies only near the band
edges. \ We take  the density of states as a semicircle, $N(\varepsilon
)=b^{3}\int \frac{d^{3}p}{\left( 2\pi \right) ^{3}}\delta (\varepsilon
-\varepsilon _{p})=\sqrt{4t^{2}-\varepsilon ^{2}}/2\pi t^{2}$ with parameter
$t$ chosen to match the band-edge density of states, $t=\left( 4\pi \right)
^{2/3}/2mb^{2}$ ($t=0.75eV$ and $1.5eV$ for the GaAs heavy and light hole bands).
This choice of cutoff corresponds to a Bethe lattice in
infinite dimensions; the crucial point is that it has the physically correct
band edge density of states, this is the only aspect
important for our work.  Then  $g_{0}$ obeys
the equation ${\bf g}_{0}^{a}(\omega )={\bf g}_{0}^{b}(\omega )=\omega +\mu
-xt^{2}\langle \left( {\bf g}_{0}^{a}(\omega )+J{\bf S}\cdot {\bf \sigma }%
_{\alpha \beta }\right) ^{-1}\rangle -(1-x)t^{2}{{\bf g}_{0}^{b}(\omega )}%
^{-1}$  where the angular brackets denote averages performed in the ensemble
defined by the appropriate $Z_{loc}$.

\begin{figure}
\begin{center}
\leavevmode
\epsfxsize=7cm
\epsffile{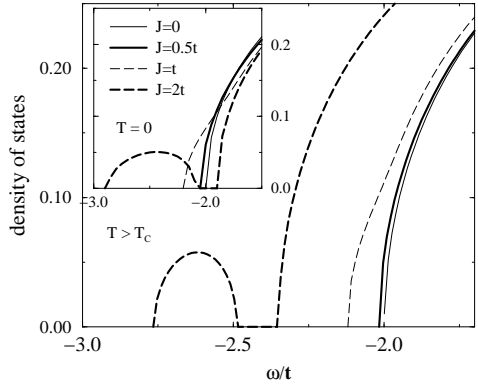}
\end{center}
\caption{{\ Main panel: density of states at $T>T_c$ for $J=0,0.5t,t$ and $2t
$. Inset: majority-spin 
density of states at $T=0$ for the same parameter values.}}
\label{dosj}
\end{figure}
\vspace{0.05in}

The solution of the equation depends crucially on $J/t$, $x$ and $T$. The
inset to Fig.~\ref{dosj} shows the majority-spin density of states
corresponding to the $T=0$ ferromagnetic state. For small $J$ we see the
expected shift proportional to $xJ$. For  $J>J_{c}=t$ an impurity band
centered at $\sim -J$ and containing $x$ states is seen to split off from
the main band. The DMFT $J_{c}$ is in good numerical agreement with the
results of \cite{Sanvito01}; this and the obviously correct qualitative
behavior confirms its reliability in the experimentally relevant regime.

As the temperature is increased, the spins disorder and eventually the
magnetic transition temperature is reached. Above this temperature, $g_{0}$
is spin-independent. The main panel of Fig.~\ref{dosj} shows the density of
states for $T>T_{c}$. For $xJ^{2}/(t^{2}-J^{2})<<1$ there is a small spin
independent band offset of size $xJ^{2}/(t^{2}-J^{2})$. For $J>t$ an
impurity band forms, corresponding to carriers locally parallel to Mn
spins.

The ferromagnetic transition temperature $T_{c}$ may be obtained by linearizing
the equation in the magnetic part of $g_{0}$, leading to an implicit equation
for $T_{c}$.
\begin{equation}
1=\sum_{n}\frac{-2t^{2}(xJ)^{2}/3}{(g_{0}^{2}-x^{2}J^{2})^{2}
\left(1-t^{2}/g_0^{2}\right)
-xJ^{2}t^{2}(5/3-J^{2}/g_{0}^{2})}  \label{Tcgen}
\end{equation}
where temperature is contained in the Matsubara sum over the frequency $%
\omega _{n}$ on which $g_0$ depends.
\begin{figure}
\begin{center}
\leavevmode
\epsfxsize=7cm
\epsffile{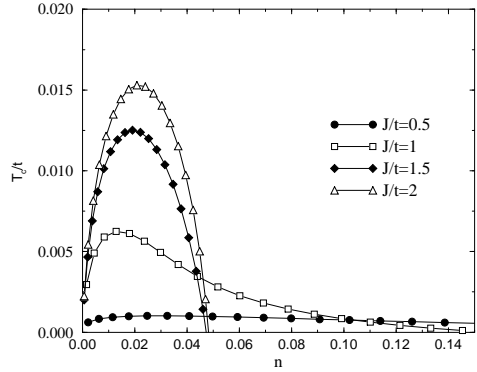}
\end{center}
\caption{{\ Calculated transition temperature vs
carrier concentration for $x=0.05$ and various J values
as shown}}
\label{Tcn}
\end{figure}

Figure.~\ref{Tcn} shows the electron density dependence of the magnetic
transition temperature for $J=0.5t$ (less than the critical value for
impurity band formation), $J=t$ (the critical value for impurity band
formation) and $J=1.5t$ and $2t$ (where the impurity band is well formed).
The striking feature, evident in all three curves, is the non-monotonic
behavior of the transition temperature. This has different origins in
different regimes. For $xJ^{2}/\left( t^{2}-J^{2}\right) <1$ an analytic
solution for $T_{c}$ may be obtained. The details will be presented
elsewhere; one result is that the density $n^{\max }$ at which $T_{c}$ is
maximized is $n^{\max }=\frac{1}{\pi }(2-\sqrt{3+2J^{2}/t^{2}-J^{4}/t^{4}}%
)^{3/2}+{\cal O}(xJ^{2}/\left( t^{2}-J^{2}\right) )\bar{)}\approx 0.04-{\cal %
O(}J/t)$ $\allowbreak $. Thus in this limit  the $T_{c}$ maximum is a
consequence of structure in the underlying electronic susceptibility. The
precise position depends on the  cutoff, but is very low. For $J>t$
the physics is dominated by the spin-polarized impurity band. In this limit $%
T_{c}$ is controlled by the delocalization energy in the impurity band, and
is therefore maximized when the band is half filled. In a filled impurity
band ($n=x$) no low energy hopping processes are allowed in a ferromagnetic
state whereas in an antiferromagnetic state hopping is allowed with
amplitude $x^{1/2}t/\Delta $ \ where $\Delta \sim J$ is the gap between the
impurity and conduction band. This physics implies that very near the filled
impurity band limit, the ground state is antiferromagnetic. As $\Delta $
increases the window of antiferromagnetism decreases.
\begin{figure}
\begin{center}
\leavevmode
\epsfxsize=7cm
\epsffile{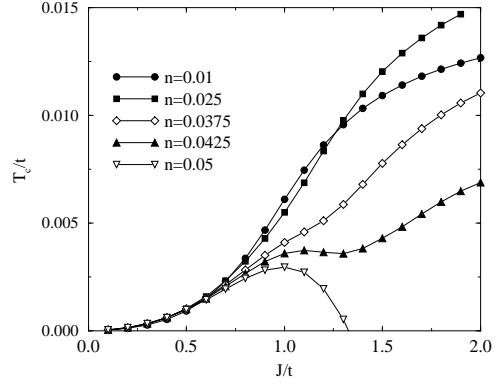}
\end{center}
\vspace{0.05in}
\vspace{0.05in}
\caption{{ Calculated $T_c$
as a function of the local exchange coupling $J$ for a fixed
value ($x=0.05$) of the Mn  concentration and for different values of the hole
density per Mn ion ($n_h = 0.2,0.5,0.75,0.85,1.0$ as shown).}}
\label{TcJ}
\end{figure}

Fig.~\ref{TcJ} shows the magnetic transition temperature as a function of
magnetic coupling $J$ for different hole densities ranging from $n_h=$
0.1/Mn to 1/Mn ( in our conventions, $n=xn_h$) The collapse in $T_{c}$ for the
filled band is evident. The physically evident decrease of $T_{c}$ at very
large $J$ due to decrease of impurity state Bohr radius is not captured by
our model, so we expect that for $J >  2t$ our calculation overestimates
$T_c$.

Fig 4 shows the dependence of $T_{c}$ on Mn concentration $x$ for
$J=t$, a value of the order of the
LSDA estimate. We see that simultaneous increases
in the Mn concentration (by say a factor of $2$) and the density
(by say a factor of $4$) should increase $T_c$ by more than a factor of two.
\begin{figure}
\begin{center}
\leavevmode
\epsfxsize=7cm
\epsffile{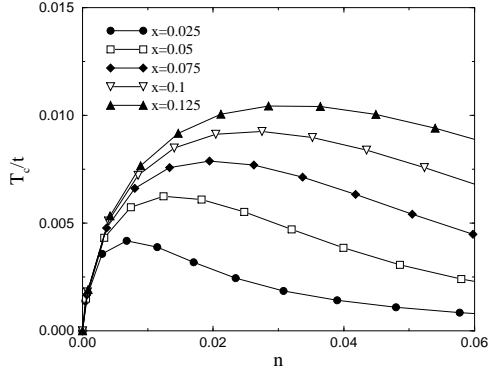}
\end{center}
\caption{{ Variation of $T_c$ with $n$ at various Mn concentration $x$
and $J=1$.
 }}
\label{tcnj1}
\end{figure}
\vspace{0.05in}

We compare our results to the predictions of other means
of calculation. The 'mean field theory' \cite{Dietl00} applied to our model
predicts $T_{c}=xn^{1/3}J^{2}/t$ at all $n,J$. In the limit $xJ^{2}/\left(
t^{2}-J^{2}\right) <1$ and $n^{\min }<n<n^{\max }$ our analytic solution of
the equations yields the mean field result but we find deviations either as $%
J$ approaches $t$, or when $n$ exceeds $n^{\max }$ or in the extremely low
density limit $n<\left( xJ^{2}/t\right)^{3/2}$ (not visible in the plots
shown here) where $T_{c}\sim xn$. An alternative approach to $T_{c}$
involves spin-wave excitations \cite{Konig00,Nagaev96}.
In classical high-spin magnets at $T\sim
T_{c}$ spin waves are excited throughout the Brillouin zone and $T_{c}$
occurs when the number of excitations (set by T divided by a typical magnon
energy) is large enough. The present theory may be thought of as a
calculation of a typical (i.e. averaged over the zone) magnon energy (which
is itself determined by the changes in electronic energy due to spin
disorder) on the assumption that the spin wave excitations have no
particular spatial structure. In $d=3$ for $T$ near $T_{c}$ this is correct
except for small amplitude critical fluctuations of no particular energetic
significance. More importantly, our calculation provides detailed
access to the experimentally relevant intermediate J,n,x regimes.

We now briefly discuss numerical estimates of $T_{c}$ for GaAs:Mn. The band
theory estimates \cite{Sanvito01} $J=1$eV and $t=0.75eV$ (corresponding to
the heavy hole mass) along with the typical density $n_{h}=$0.1/Mn, implies
a $J/t\approx 1.3$ and a $T_{c}\approx 80K$ for $x=0.05$ whereas the light
hole mass ($t=1.5eV$) implies $J/t\approx 0.75$ and a $T_{c}\approx 50K$.
These $T_{c}$ values are for the single band model; the contributions of the
two bands add, leading to a $T_{c}^{phys}\approx 130K$. Although our
theoretical estimates agree well with the experimental $T_{c}$\cite
{Ohno98,Ohno96}, this agreement should not be taken too seriously in view of
the simplifying approximations of our model. Interestingly, for the LSDA $J$
the heavy hole band makes a higher contribution to $T_{c}$. Increase of $n$
by about $50\%$ will increase $T_{c}$ by a similar amount for the heavy hole
band but less for the light hole band. Increases in $J$ if it can be managed
will also increase $T_{c}$ (although much
less rapidly than the quadratic
dependence predicted by the mean field theory). The most promising route to
a higher temperature ferromagnet is predicted to be a simultaneous increase
in $x$ to a value of order $0.1$ and $n$ to about $0.02$ or about 0.2/Mn$.$

In summary, we have presented a theory of the magnetic semiconductors which
can handle both the weak coupling limit ($xJ$ less than Fermi energy $E_{F}$%
) and the intermediate coupling regime ($J>t$ but not too large). This method
correctly treats the physically relevant situation in which the carriers are
constrained to be locally parallel to the Mn spins and allows, for example,
calculation of the resistivity and optical conductivity. Discussion of these
quantities will be given elsewhere.

This work is supported by NSF-DMR-MRSEC-0080008 (AC), US-ONR and DARPA (SDS)
and NSF-DMR-00081075 (AJM).

\end{document}